\documentstyle[sprocl,psfig]{article}

\def\Journal#1#2#3#4{{#1} {\bf #2}, #3 (#4)}

\def\NPB{{\em Nucl. Phys.} B}
\def\PLB{{\em Phys. Lett.}  B}
\def\PRL{\em Phys. Rev. Lett.}
\def\PRD{{\em Phys. Rev.} D}
\def\ZPC{{\em Z. Phys.} C}
\def\PTP{\em Prog. Theor. Phys.}
\def\EPJ{\em {Eur. Phys. J.} C}

\def\be{\begin{equation}}
\def\ee{\end{equation}}
\def\bea{\begin{eqnarray}}
\def\eea{\end{eqnarray}}

\newcommand{\etal}{{\it et al.\/}}

\newcommand{\ol}[1]{\overline{#1}}

\newcommand{\sign}{\mathop{\rm sign}\nolimits}
\newcommand{\br}{\mathop{\rm B}\nolimits}   
\newcommand{\gsim}{\mbox{ \raisebox{-1.0ex}{$\stackrel{\textstyle >}
{\textstyle \sim}$ }}}
\newcommand{\lsim}{\mbox{ \raisebox{-1.0ex}{$\stackrel{\textstyle <}
{\textstyle \sim}$ }}}

\newcommand{\tanb}{\tan\beta}
\newcommand{\bb}{$B^0$--$\ol{B}^0$}

\newcommand{\ek}{\epsilon_K}
\newcommand{\bsg}{b\rightarrow s\, \gamma}
\newcommand{\brbsg}{\br(\bsg)}
\newcommand{\bsll}{b\rightarrow s\, l^+\, l^-}
\newcommand{\kpnn}{K\rightarrow \pi\, \nu\, \ol{\nu}}
\newcommand{\klpnn}{K_L\rightarrow \pi^0\, \nu\, \ol{\nu}}
\newcommand{\kppnn}{K^+\rightarrow \pi^+\, \nu\, \ol{\nu}}
\newcommand{\dmbd}{\Delta M_{B_d}}

\newcommand{\rxd}{\dmbd\slash\dmbd^{\rm SM}}

\newcommand{\amu}{a_\mu^{\rm SUSY}}

\begin{document}
\begin{flushright}
  \begin{tabular}[t]{l} 
  KEK-TH-657\\
  October 1999
 \end{tabular}
 \end{flushright}
\vspace*{0.5cm}

\title{Indirect Search for Supersymmetry
\footnote{
Lecture given at International Symposium on Supersymmetry,
Supergravity, and Superstring, June 23 -26, 1999, Seoul, Korea. 
}}

\author{YASUHIRO OKADA}

\address{Institute for Particle and Nuclear Studies, KEK,\\
Oho 1-1, Tsukuba 305-0801, Japan\\E-mail: yasuhiro.okada@kek.jp} 




\maketitle\abstracts{Effects of supersymmetric particles 
may appear in various low energy experiments through loop
diagrams. We discuss various flavor changing neutral current 
processes, the muon anomalous magnetic moment and lepton flavor 
violation in the context of the supergravity model. In particular
the \bb\ mixing and the muon anomalous magnetic moment 
are revisited taking into account the recent Higgs boson search
result. We also consider $\mu^+ \rightarrow e^+ \gamma$
and  $\mu^+ \rightarrow e^+ e^+ e^-$ processes with polarized
muons. We calculate the P-odd and T-odd asymmetries in these
processes for SU(5) and SO(10) supersymmetric grand unified 
theories and show that these asymmetries are useful to 
distinguish different models. }

\section{Introduction}
Unified theories based on supersymmetry (SUSY) 
have been studied as one of promising candidates beyond the 
Standard Model (SM) since early 1980's because this symmetry 
could be a solution of hierarchy problem
in the SM. Motivation for SUSY unification was strengthened
when precisely determined values of three gauge coupling constants
turned out to be consistent with the SU(5) supersymmetric 
grand unified theory (SUSY GUT). Experimental verification
of SUSY is, therefore, one of the most important issues 
of current high energy physics. 

In order to explore SUSY indirect search experiments are 
important in addition to direct search for SUSY particles
at collider experiments. There are variety of possibilities  
that SUSY effects can appear in low energy experiments. 
SUSY can affect flavor changing neutral current
(FCNC) processes and CP violation in B and K meson decays.
Processes like proton decay, lepton flavor violation (LFV) such as 
$\mu \rightarrow e \gamma$ and neutron and electron electric 
dipole moments (EDM) are important because these are either 
forbidden or strongly suppressed within the SM.

In the context of SUSY models flavor physics has important 
implications. Because the squark and the slepton
mass matrices become new sources of flavor mixings 
generic mass matrices would induce too large FCNC 
and LFV effects if the superpartners' masses are in a few-hundred-GeV
region. For example, if we assume that the SUSY contribution to the  
$K^0 - \bar{K}^0$ mixing is suppressed because of the cancellation
among the squark contributions of different generations, the squarks
with the same $SU(3)\times SU(2)\times U(1)$ quantum numbers should
be highly degenerate in masses. When the squark mixing angle is 
in a similar magnitude to the Cabibbo angle the requirement on 
degeneracy becomes as
\begin{equation}
\frac{\Delta m_{\tilde{q}}^2}{ m_{\tilde{q}}^2}\lsim 10^{-2}
\left(\frac{ m_{\tilde{q}}}{100 GeV}\right)
\end{equation}
for at least the first and second generation squarks.
Similarly, the $\mu^+ \rightarrow e^+ \gamma$ process
puts a strong constraint on the flavor off-diagonal terms
for the slepton mass matrices which is roughly given by
\begin{equation}
\frac{\Delta m_{\tilde{\mu}\tilde{e}}^2}{ m_{\tilde{l}}^2}\lsim 10^{-3}
\left(\frac{ m_{\tilde{l}}}{100 GeV}\right)^2 .
\end{equation}

In the minimal supergravity model these flavor problems are avoided
by taking SUSY soft-breaking terms as flavor-blind structure. 
The scalar mass terms are assumed to be common for
all scalar fields at the Planck  scale and therefore there are no 
FCNC effects nor LFV  from the squark and slepton sectors at this scale. 
The physical squark and slepton masses are determined taking account 
of the renormalization effects from the Planck to the weak scale. 
This induces sizable SUSY contributions to various FCNC processes 
in the B and K decays. Detail calculations have been done 
for processes such as \bb\ mixing, $\ek$, $\bsll$ and $\kpnn$ 
in the minimal supersymmetric standard model (MSSM) and 
as well as in the minimal supergravity model.\cite{FCNC,GOS} 
Also if there is interaction which breaks lepton flavor conservation
between the Planck and the weak scales, LFV effects can be induced
through the slepton mass matrices. An important example of this kind
is SUSY GUT models where interaction at the GUT scale induces
flavor mixing in the lepton/slepton sector as well as in the 
quark/squark sector.\cite{BH}

In this talk we discuss two recent works on flavor physics in 
the supergravity model. The first topics is an update of FCNC 
processes in B and K decays in the supergravity model.\cite{GOS2} 
In this work we took into account recent improvement of Higgs boson search 
at LEP II which turns out to be very important to constrain SUSY
parameter space, especially for small $tan{\beta}$ region. We also 
calculate SUSY contributions to the anomalous magnetic moment of 
muon in this model. We show that the on-going experiment
at BNL will put a strong constraint for large $tan{\beta}$ region.   

The second topics is LFV process in the SUSY GUT models.\cite{OOS,OOS2} 
Here we consider $\mu \rightarrow e \gamma$ and $\mu \rightarrow 3 e$  
processes with polarized muons. We show that parity-odd (P-odd)
and time-reversal-odd (T-odd) asymmetries
are useful to distinguish various sources of LFV interactions.
In particular SU(5) and SO(10) SUSY GUT have different features
of these asymmetries. 

\section{Update of FCNC Processes and Muon Anomalous Magnetic Moment 
in the Supergravity Model}
\subsection{\bb\ mixing in the Supergravity Model}
In the minimal SM various FCNC processes and CP violation in B and K decays
are determined by the Cabibbo-Kobayashi-Maskawa (CKM) matrix. The constraints
on the parameters of the CKM matrix can be conveniently expressed
in terms of the unitarity triangle. With CP violation at B factory as well as
rare K decay experiments we will be able to check consistency of the 
unitarity triangle and at the same time search for effects of physics 
beyond the SM. In order to distinguish possible new physics effects it 
is important to identify how various models can modify the SM predictions.
 
Although general SUSY models can change the lengths and the angles of the 
unitarity triangle  in variety ways, the supergravity model predicts a 
specific pattern of the deviation from the SM. 
Namely, we can show that the SUSY loop contributions to FCNC amplitudes 
approximately have the same dependence on the CKM elements as the SM 
contributions. In particular, if we assume that there are no CP 
violating phases from SUSY breaking sectors, the complex phase of 
the $B^0 - \bar{B^0}$ mixing amplitude does not change even if we 
take into account the SUSY and the charged Higgs loop contributions. 
In terms of the unitarity triangle this means 
that the angle measurements through CP asymmetry in B decays determine 
the CKM matrix elements as in the SM case. On the other hand the length 
of the unitarity triangle determined from $\Delta M_B$ and $\epsilon_K$ 
can be modified. The case with supersymmetric CP phases was also 
studied within the minimal supergravity model and it was shown that 
effects of new CP phases on the $B^0 - \bar{B^0}$ mixing amplitude 
and the CP asymmetry in the $\bsg$ process are small once 
constraints from neutron and electron EDMs 
are included.\cite{Nihei,GKNOS}. 

We present the result of numerical calculation of 
$\Delta M_B$  in the supergravity model. We also calculate
the branching ratios of $b \rightarrow s\gamma$ in this model 
and this is used as a constraint on SUSY parameter space.
In the calculation we have used updated 
results of various SUSY search experiments at LEP II and Tevatron
as well as the next-to-leading QCD corrections in the calculation
of the $b \rightarrow s\gamma$ branching ratio. Most important
change in the experimental constraints for the last one year 
comes from improvement of SUSY Higgs boson search at LEP II.

It is well known that there is a strict upper bound on the lightest
CP-even Higgs boson mass in the MSSM.\cite{Vloop} 
Because the Higgs search in the LEP~II experiment has reached to the
sensitivity of 100 GeV for the SM Higgs boson and about 90 GeV for
the case of the lightest CP-even Higgs boson in the MSSM, a meaningful
amount of the SUSY parameter space is already excluded.\cite{LEPhiggs}

We first present the magnitude of $\dmbd$ in the supergravity
model. In Fig. 1 we show $\rxd$ as a function of the lightest Higgs
boson mass for $\tanb$ = 2, where
$\dmbd^{\rm SM}$ denotes the SM value.
\begin{figure}
\begin{center}
\mbox{\psfig{figure=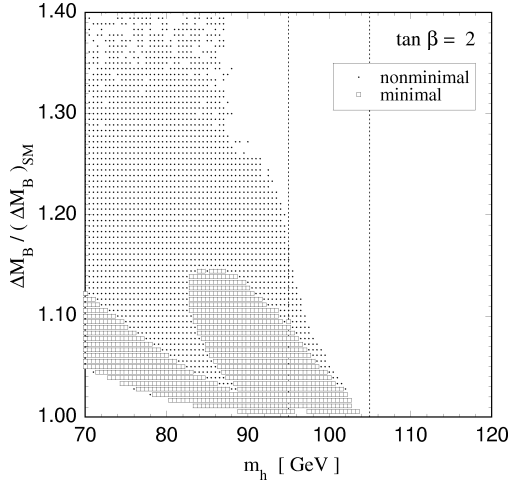,width=2.6in,angle=0}}
\end{center}
\caption{$\Delta M_{B_d}$ normalized by the SM value for $\tan{\beta}=2$.
The square(dot) points correspond to the minimal (enlarged) parameter space of
the supergravity model.
\label{fig:fig1}}
\end{figure} 
As input parameters we take
$m_t^{\rm pole} = 175$ GeV,
$m_b^{\rm pole} = 4.8$ GeV,
$\alpha_s(m_Z) = 0.119$.
For  the CKM matrix elements, we take $V_{us} = 0.2196$, 
$V_{cb} = 0.0395$, $|V_{ub}/V_{cb}| = 0.08$ and the CP violating 
phase in the CKM matrix as $\delta_{13}=\pi/2$. 
In the supergravity model $\rxd$ 
is the almost independent of the CKM parameters,
so that the following results do not change very much
if we take different values of the CKM matrix elements.
We have calculated the SUSY particle spectrum based on
two different assumptions on the initial conditions of RGE.\cite{GOS} 
The minimal case corresponds to the minimal supergravity 
where all scalar fields have a common SUSY breaking mass at the GUT scale. 
In the second case shown as ``nonminimal'' in the figures we enlarge 
the SUSY parameter space by relaxing the initial conditions for the 
SUSY breaking parameters, namely all squarks and sleptons have a 
common SUSY breaking mass whereas an independent
SUSY breaking parameter is assigned for Higgs fields. From this 
figure we see that the deviation from the SM of $\dmbd$ is reduced 
to $+15$\% for the nonminimal case if we require that the lightest 
CP-even Higgs mass should be larger than 95 GeV for $\tanb = 2$. 
If the mass bound is raised to 105 GeV no allowed parameter space
remains for $\tanb = 2$.

In Fig. 2 the allowed ranges of $\rxd$ are shown for
several values of $\tanb$ with the present constraint of the SUSY Higgs
boson search and Fig.3 corresponds to the case where
$m_h > 105$ GeV. 
\begin{figure} 
\begin{center}
\mbox{\psfig{figure= 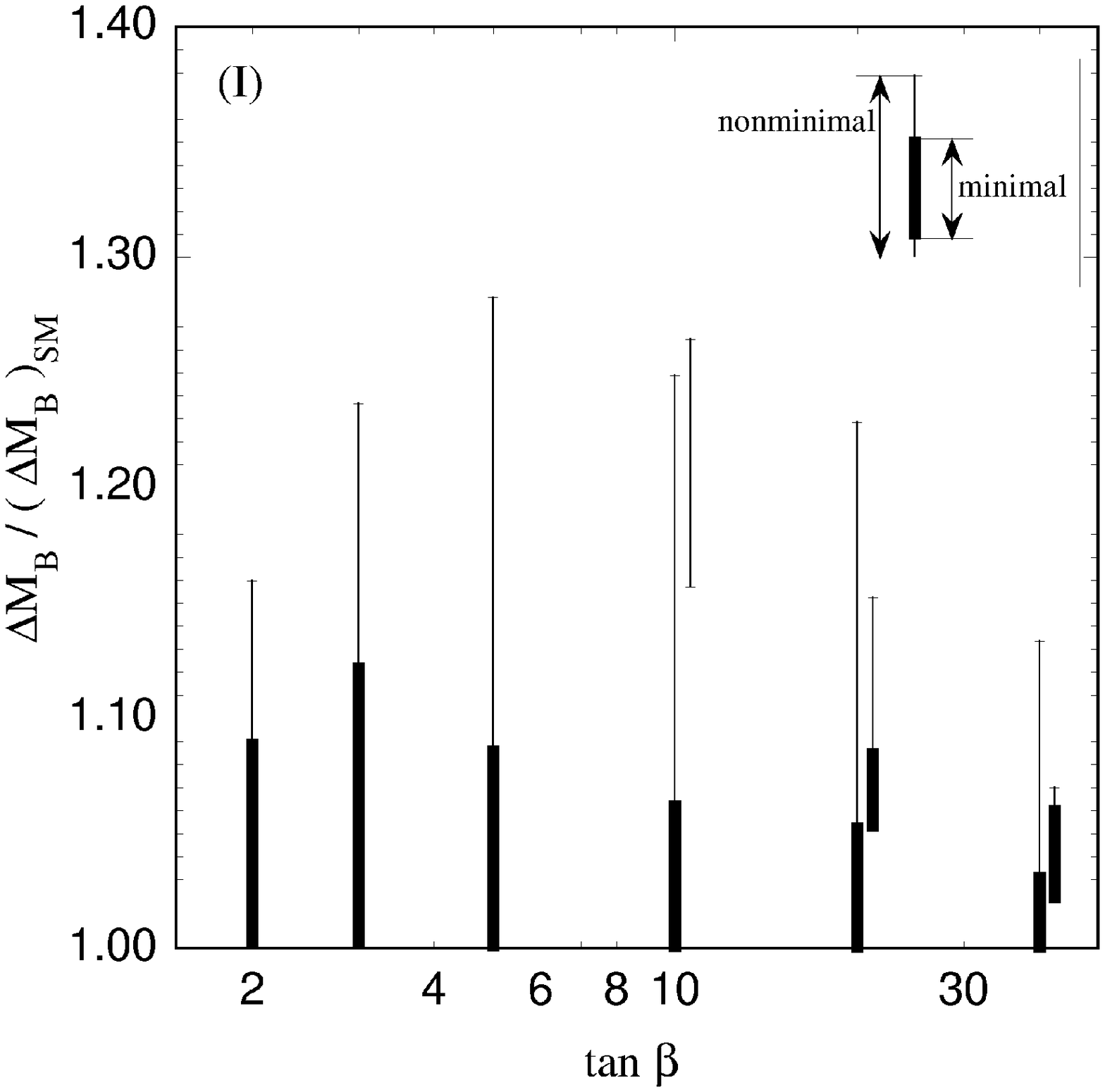,width=2.6in,angle=0}}
\end{center}
\caption{  Allowed ranges of $\rxd$for several values of $\tanb$ with 
  the present constraint of the Higgs boson mass.
  Two lines are shown according to the sign of the $\bsg$amplitude 
  for $\tanb=10,\,20,\,40$. The left (right) lines correspond the case
  where the sign of the $\bsg$ amplitude is same (opposite) as that in
  the SM. 
\label{fig:fig2}}
\end{figure} 
\begin{figure} 
\begin{center}
\mbox{\psfig{figure= 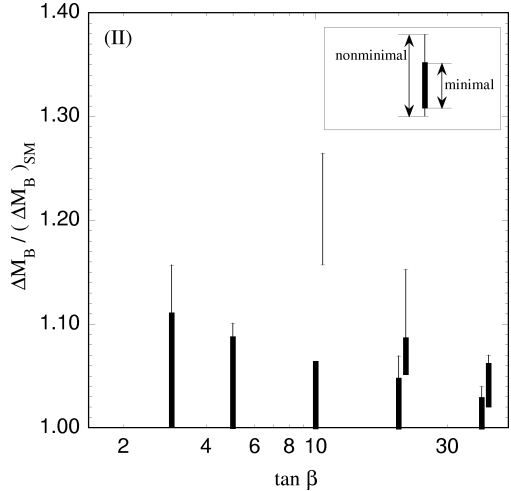,width=2.6in,angle=0}}
\end{center}
\caption{ The same figure as Fig.2 for the case that $m_h > 105$ GeV.
\label{fig:fig3}}
\end{figure} 
We take into account both $e^+e^- \rightarrow
Z h$ and $e^+e^- \rightarrow A h$ processes to constrain the
SUSY parameter space from the SUSY Higgs boson search.\cite{GOS2}
For $\tanb \gsim 10$ there appears a separate region where the sign of
the $\bsg$  amplitude is opposite to that of the SM amplitude.
In this region the SUSY contribution to the $\bsg$ amplitude 
is large and opposite in sign to the SM amplitude. 
Two cases are distinguished according to the sign of the $\bsg$ amplitude
for $\tanb = 10,\,20,\,40$. 
We can see that the present constraint allows the deviation of the SM
for the \bb mixing up to 12\% for the minimal case and 30\% for the
nonminimal case.
If the Higgs boson bound is raised to 105 GeV, the allowed deviation
from the SM becomes less than 15\% except for the small parameter space
where the sign of the $\bsg$ amplitude becomes opposite to the SM case.
Note that in such parameter space the $\bsll$ branching ratio
is expected to be enhanced.\cite{FCNC,GKNOS} 
Deviations from the SM of $\ek$, $\klpnn$, and $\kppnn$ are also
studied.\cite{GOS2} If $\ek$ is normalized by the SM value 
the deviation from the SM is essentially the same as that of $\dmbd$.
On the other hand, the both $\kpnn$ branching ratios are reduced 
up to 5\% for the minimal case and 15\% for the nonminimal case with 
the present Higgs bounds and further reduced to 7\% for $m_h > 105$ GeV
in the nonminimal case.  

These deviations may be evident in future when B factory experiments 
provide additional information on the CKM parameters. It is expected 
that the one of the three angles of the unitarity triangle is determined 
well through the $B \rightarrow J/\psi K_S$ mode. Then assuming the SM,
one more physical observable can determine the CKM parameters or 
$(\rho,\eta)$ in the Wolfenstein parameterization.
New physics effects may appear as inconsistency in the determination 
of these parameters from different inputs. 
For example, the $\rho$ and $\eta$ parameters 
determined from CP asymmetry of B decay in other modes, $\frac
{\Delta M_{B_s}}{\Delta M_{B_d}}$ and $|V_{ub}|$ can be considerably
different from those determined through  $\Delta M_{B_d}$ 
$\epsilon_K$ and B($K\rightarrow \pi \nu \bar{\nu})$ because
$\Delta M_{B_d}$ $\epsilon_K$ are enhanced and 
B($K\rightarrow \pi \nu \bar{\nu})$ is reduced but
$\frac{\Delta M_{B_s}}{\Delta M_{B_d}}$ and $|V_{ub}|$ are
essentially independent of SUSY contributions in this model.
The pattern of these deviations from the SM will be a key 
to distinguish various new physics effects.

\subsection{Muon anomalous Magnetic Moment in the Supergravity Model} 
The muon anomalous magnetic moment is a very precisely measured quantity.
The present experimental value of $a_\mu = (g-2)_\mu/2$ is
$  a_\mu^{\rm exp} = 11659235.0 (73.0) \times 10^{-10}$.\cite{PDG,BNL}
This is also one of most accurately calculated 
quantities within the SM. The SM prediction is 
$ a_\mu^{\rm SM} = 11659109.6 (6.7) \times 10^{-10}$
where the error in the SM value is dominated by the hadronic
contribution of the vacuum polarization diagram.\cite{Marciano,hadronic}
Because the QED corrections are very precisely known
the experimental accuracy is now approaching to the level of detecting 
possible new physics contributions at the electroweak scale. 
Combining the above two values we can derive a possible new physics
contribution to $a_\mu$ as
\begin{eqnarray}
  a_\mu^{\rm exp} - a_\mu^{\rm SM} &=&
  ( 75.5 \pm 73.3 ) \times 10^{-10}
~.
\end{eqnarray}
The current BNL experiment is aiming to improve $a_\mu$ by a factor of
20 and the first result is reported as $a_\mu = 1165925(15) \times
10^{-9}$.\cite{BNL}

In the context of SUSY models it is known that there is sizable
contributions to $a_\mu$ ($\amu$) from the loop diagrams with
sneutrino and chargino and with charged slepton and neutralino.
The SUSY effects are particularly important for large $\tanb$
region. The $\amu$ was calculated in the MSSM \cite{g-2mssm}
as well as in the supergravity model.\cite{g-2sugra,CN}  

We calculated the SUSY contribution to $a_\mu$ ($\amu$) in the
supergravity model discussed above. We require the radiative 
electroweak symmetry breaking condition and the various 
phenomenological constraints as before.

We first present $\amu$ in the minimal supergravity for 
$\tanb$=10 as a function of $\brbsg$ in Fig. 4.
\begin{figure} 
\begin{center}
\mbox{\psfig{figure= 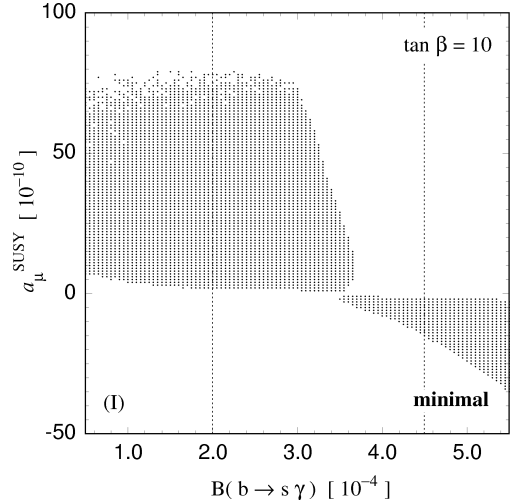,width=2.6in,angle=0}}
\end{center}
\caption{$\amu$ in the supergravity model for $\tan{\beta}=10$.
The square(dot) points correspond to the minimal (enlarged) parameter 
space of the supergravity model.
\label{fig:fig4}}
\end{figure} 
In this figure the present bound from the Higgs boson search is 
applied. The magnitude of $\amu$ becomes large for
large $\tanb$. The enhancement of $\amu$ for large $\tanb$ comes 
from the fact that $\amu$ is dominated with the sneutrino-chargino 
loop diagram, which contains a contribution proportional to $\mu\tanb$.
As a result, SUSY contributions to both $\bsg$ amplitude and
$\amu$ are correlated with $\sign(\mu)$.
We can see that $\amu$ becomes positive (negative) according
to the suppression (enhancement ) of $\brbsg$.
This correlation was pointed out in the literature.\cite{CN}

The predicted range of $\amu$ are shown for several values
of $\tanb$ with the present constraint on the Higgs search and with
$m_h > 105$ GeV for the minimal and the nonminimal cases in Fig. 5
and Fig. 6. 
\begin{figure} 
\begin{center}
\mbox{\psfig{figure= 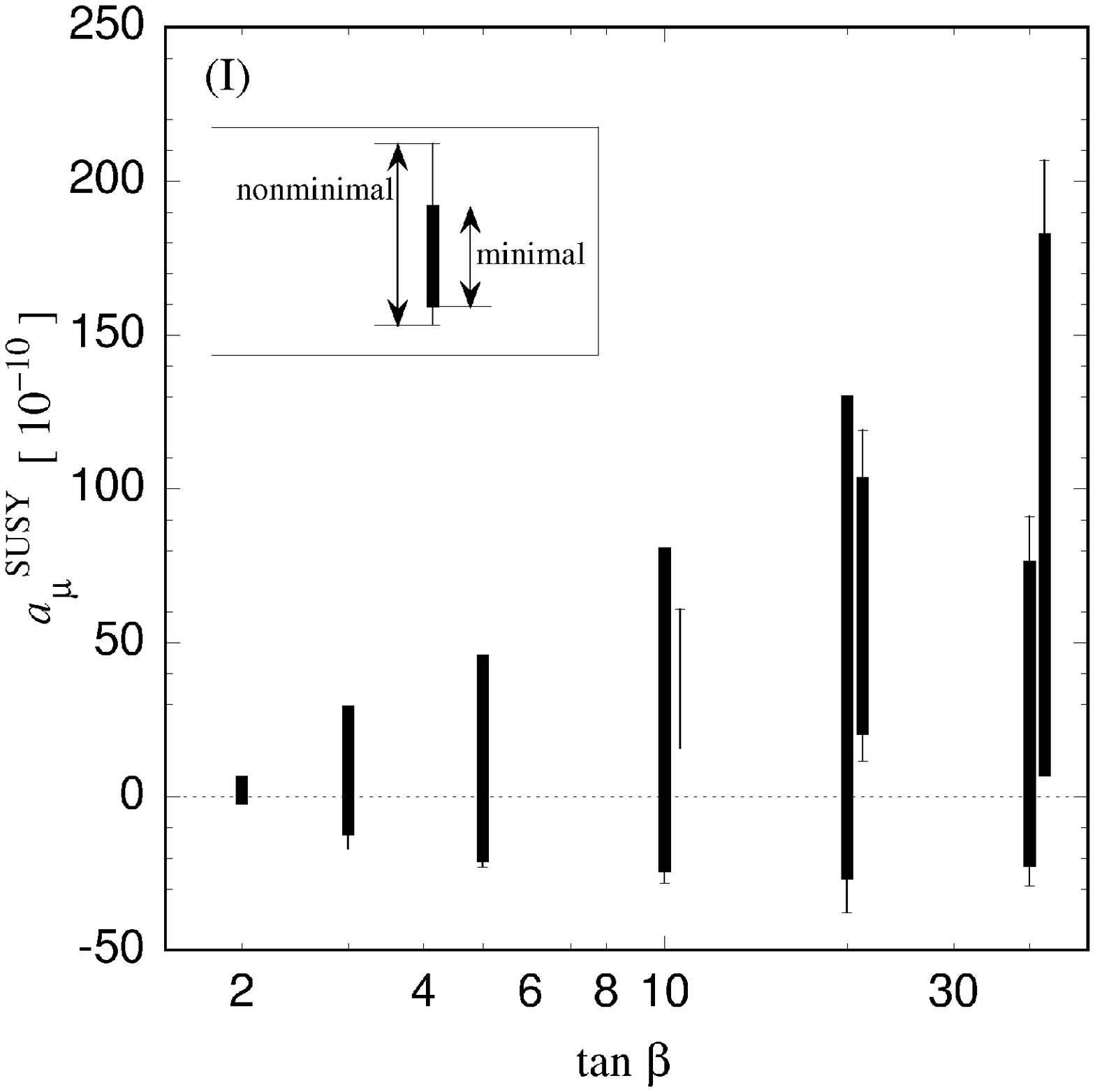,width=2.6in,angle=0}}
\end{center}
\caption{ Allowed ranges of $\amu$for several values of $\tanb$ with 
  the present constraint of the Higgs boson mass.
  Two lines are shown according to the sign of the $\bsg$amplitude 
  for $\tanb=10,\,20,\,40$. The left (right) lines correspond the case
  where the sign of the $\bsg$ amplitude is same (opposite) as that in
  the SM.
\label{fig:fig5}}
\end{figure} 
\begin{figure} 
\begin{center}
\mbox{\psfig{figure= 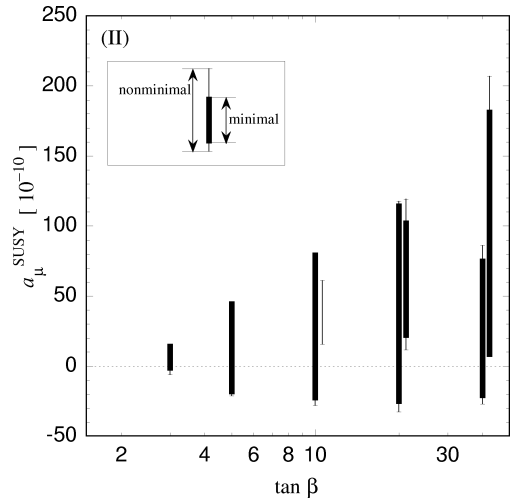,width=2.6in,angle=0}}
\end{center}
\caption{ The same figure as Fig.5 for the case that $m_h > 105$ GeV.
\label{fig:fig6}}
\end{figure} 
Two cases according to the sign of the $\bsg$  amplitude 
are shown separately for $\tanb$ =$10,\, 20,\, 40$. This figures show 
that the muon anomalous magnetic moment is indeed expected to become 
very powerful to constrain the SUSY parameter space in near future.
We see that even for $\tanb$ =$ 5$ the deviation is quite sizable
considering future improvements on the $a_\mu$ measurement.

\section{$\mu^+ \rightarrow e^+ \gamma$
and $\mu^+ \rightarrow e^+ e^+ e^-$ with Polarized Muons in SUSY GUT}
\subsection{P-odd and T-odd asymmetries in $\mu^+ \rightarrow e^+ \gamma$
and $\mu^+ \rightarrow e^+ e^+ e^-$ processes}

LFV processes such as $\mu^+ \rightarrow e^+ \gamma$, 
$\mu^+ \rightarrow e^+ e^+ e^-$ and $\mu^-  - e^-$ conversion in atoms
are another interesting possibility to search for SUSY effects through 
flavor physics.
The current experimental upper bounds on these processes are
$B(\mu^{+} \rightarrow e^{+} \gamma) \leq 1.2 \times10^{-11}$,\cite{MEGA}
$B(\mu^{+} \rightarrow e^{+}e^{+}e^{-}) \leq 1.0\times 10^{-12}$,\cite{mu3eex} 
and $\sigma(\mu^{-}$Ti$ \rightarrow e^{-}$Ti$)/
\sigma(\mu^{-}$Ti$ \rightarrow capture) 
\leq 6.1\times10^{-13}$.\cite{conversionex}
Recently there are considerable interests on these processes because
predicted branching ratios turn out to be close to the upper bounds in 
the SUSY GUT.\cite{BH}

As discussed before no LFV is generated at the Planck scale
in the context of the minimal supergravity model. In the SUSY GUT 
scenario, however, the LFV can be induced through renormalization 
effects on the slepton mass matrix because GUT interaction breaks 
lepton flavor conservation. In the minimal SUSY SU(5) GUT, the effect 
of the large top Yukawa coupling constant results in the LFV in the 
right-handed slepton sector.\cite{BH,BHS,97Hisano} 
On the other hand in the SO(10) model both left- and right-handed 
sleptons induce LFV and the predicted branching ratio can be larger 
by two order of magnitudes.\cite{BHS} A similar enhancement is 
seen for SU(5) model with large $\tanb$ when we take into acccount 
higher dimensinal operators for the Yukawa couplings at the GUT scale 
to explain realistic fermion mass relations.\cite{98Hisano} 

In this section we discuss $\mu^{+}\rightarrow e^{+}\gamma$ and 
$\mu^{+} \rightarrow e^{+}e^{+}e^{-}$ with polarized muons.
Experimentally highly polarized $\mu^+$ beam is available
because muons from decay of pions stopped at the target surface 
is initally 100\% polarized opposite to the muon momentum direction.
This muon beam is called surface muon. It was pointed out that
polarized muon is useful to suppress background processes
for $\mu^{+}\rightarrow e^{+}\gamma$ search because main
background processes have particular angular 
distributions.\cite{96Kuno} 

The muon polarization is also useful
in identifying the nature of LFV interactions. 
Using angular distributions of final particles we can define
asymmetries with respect to the inital muon polarization 
direction. For $\mu^{+}\rightarrow e^{+}\gamma$ we define
a P-odd asymmetry which distinguishes between
$\mu^{+}\rightarrow e_L^{+}\gamma$ and 
$\mu^{+}\rightarrow e_R^{+}\gamma$. For the 
$\mu^{+} \rightarrow e^{+}e^{+}e^{-}$ process we can define 
two P-odd asymmetries and one T-odd asymmetry. From these 
asymmetries we can obtain information on
structure of LFV interactions. In particular we calculate
branching ratios and asymmetries in the SU(5) and SO(10)   
SUSY GUT models and show that these asymmetries are useful
to distingush models. 

Using the electromagnetic gauge invariance and the Fierz 
rearrangement the effective Lagrangian for 
$\mu^{+}\rightarrow e^{+}\gamma$ and 
$\mu^{+} \rightarrow e^{+}e^{+}e^{-}$ processes can be written  
without loss of generality,\cite{OOS,OOS2}

\begin{eqnarray}
{\cal L} &=& -\frac{4G_F}{\sqrt{2}}\{  
        {m_{\mu }}{A_R}\overline{\mu_{R}}
        {{\sigma }^{\mu \nu}{e_L}{F_{\mu \nu}}}
       + {m_{\mu }}{A_L}\overline{\mu_{L}}
        {{\sigma }^{\mu \nu}{e_R}{F_{\mu \nu}}} \nonumber \\
    &&   +{g_1}(\overline{{{\mu }_R}}{e_L})
              (\overline{{e_R}}{e_L})
       + {g_2}(\overline{{{\mu }_L}}{e_R})
              (\overline{{e_L}}{e_R}) \nonumber \\
    &&   +{g_3}(\overline{{{\mu }_R}}{{\gamma }^{\mu }}{e_R})
              (\overline{{e_R}}{{\gamma }_{\mu }}{e_R})
       + {g_4}(\overline{{{\mu }_L}}{{\gamma }^{\mu }}{e_L})
              (\overline{{e_L}}{{\gamma }_{\mu }}{e_L})  \nonumber \\
    &&   +{g_5}(\overline{{{\mu }_R}}{{\gamma }^{\mu }}{e_R})
              (\overline{{e_L}}{{\gamma }_{\mu }}{e_L})
       + {g_6}(\overline{{{\mu }_L}}{{\gamma }^{\mu }}{e_L})
              (\overline{{e_R}}{{\gamma }_{\mu }}{e_R})
       +  h.c. \},
\label{eq:effective}
\end{eqnarray}
where $G_{F}$ is the Fermi coupling constant and $m_{\mu}$ is the muon mass.
The chirality projection is defined by the projection operators 
$P_{R}=\frac{1+\gamma_{5}}{2}$ and $P_{L}=\frac{1-\gamma_{5}}{2}$.
$A_{L}$($A_{R}$) is the dimensionless photon-penguin coupling 
constant which contributes to 
$\mu^{+} \rightarrow e_{L}^{+}\gamma$ ($\mu^{+} \rightarrow e_{R}^{+}\gamma$).
These couplings also induce the $\mu^{+} \rightarrow e^{+}e^{+}e^{-}$ process. 
$g_{i}$'s $\; (i=1-6)$ are dimensionless four-fermion coupling 
constants which only contribute to $\mu^{+} \rightarrow e^{+}e^{+}e^{-}$.
$A_{L,R}$ and $g_{i}\; (i=1-6)$ are generally complex numbers and 
calculated based on a particular model with LFV interactions. 

The differential branching ratio for $\mu^{+} \rightarrow e^{+}\gamma$ 
is given by: 
~
\begin{eqnarray}
\frac{dB(\mu^{+} \rightarrow e^{+}\gamma)}{d\cos\theta} &=&
                192\pi^{2}\{|A_{L}|^{2}(1+P\cos\theta)
                           +|A_{R}|^{2}(1-P\cos\theta)\}
 \\
             &=& \frac{B(\mu^{+}\rightarrow e^{+}\gamma)}{2}
                 \{1+A(\mu^{+}\rightarrow e^{+}\gamma)P\cos\theta\},
\end{eqnarray}
where the total branching ratio for $\mu^{+}\rightarrow e^{+}\gamma$ 
($B(\mu^{+}\rightarrow e^{+}\gamma)$) and the P-odd asymmetry 
($A(\mu^{+}\rightarrow e^{+}\gamma)$) are defined as
\begin{eqnarray}
B(\mu^{+}\rightarrow e^{+}\gamma) &=& 384\pi^{2}(|A_{L}|^{2}+|A_{R}|^{2}),
 \\
A(\mu^{+}\rightarrow e^{+}\gamma) &=& \frac{|A_{L}|^{2}-|A_{R}|^{2}}
{|A_{L}|^{2}+|A_{R}|^{2}},
\end{eqnarray}
and $P$ is the muon polarization.

Kinematics of the $\mu^{+} \rightarrow e^{+}e^{+}e^{-}$ process with 
a polarized muon is determined by two energy variables of decay positrons 
and two angle variables which indicate the direction of the muon 
polarization with respect to the decay plane. 
In Fig.7 we take the $z$-axis as the direction of the decay electron 
momentum ($\vec{p_{3}}$) and the $z$-$x$ plane as the decay plane. 
Polar angles $(\theta,\varphi)$ 
$(0\leq \theta \leq\pi, 0\leq \varphi <2\pi)$ indicate 
the direction of the muon polarization $\vec{P}$.
\begin{figure} 
\begin{center}
\mbox{\psfig{figure=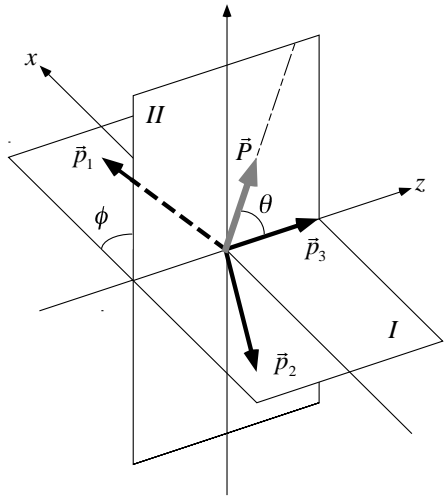 ,width=3.6in,angle=0}}
\end{center}
\caption{ Kinematics of the $\mu^{+} \rightarrow e^{+}e^{+}e^{-}$
       decay in the center-of-mass
      system of muon. The plane I represents the decay plane on which 
      the momentum vectors $\vec{p}_1$, $\vec{p}_2$, $\vec{p}_3$ lie,
      where $\vec{p}_1$ and $\vec{p}_2$ are  momenta of two $e^+$'s
      and  $\vec{p}_3$ is momentum of $e^-$ respectively.
      The plane II is the plane which the muon polarization vector
      $\vec{P}$ and $\vec{p}_3$ make.
\label{fig:fig7}}
\end{figure} 
We take a convention that the decay positron having larger energy is 
named positron 1 and the other is positron 2 and $(p_1)_x \geq 0$. 

We define two P-odd asymmetries and one T-odd asymmetry 
as asymmetries of numbers of events as follows.
\begin{eqnarray}
A_{P_1} &=& \frac{N(P_{z}>0)-N(P_{z}<0)}{N(P_{z}>0)+N(P_{z}<0)},\\
A_{P_2} &=& \frac{N(P_{x}>0)-N(P_{x}<0)}{N(P_{x}>0)+N(P_{x}<0)},\\
A_{T} &=& \frac{N(P_{y}>0)-N(P_{y}<0)}{N(P_{y}>0)+N(P_{y}<0)}.
\end{eqnarray}
Here, the $\vec{P}$ is the muon polarization vector and in the above
definition muons are assumed to be 100\% polarized. 
Using the coupling constants in the effective Lagrangian $A_{T}$ is 
expressed by 
\begin{eqnarray} 
A_{T} &=&\frac{64}{35} \frac{1}{B(\mu^{+} \rightarrow e^{+}e^{+}e^{-})}
\nonumber \\
&&\{3 Im(eA_{R}g_{4}^{*}
+eA_{L}g_{3}^{*})-2 Im(eA_{R}g_{6}^{*}+eA_{L}g_{5}^{*})\}
\end{eqnarray}
This means that in order to have a T odd asymmetry
there should be a relative phase between the photon penguin diagram
term ($A_{R}$, $A_{L}$) and four fermion terms $(g_3, g_4, g_5, g_6)$. 
As we show in the explicit numerical calculation , we need 
to introduce a CP violating phase other than the KM phase
in order to have sizable T-odd asymmetry in the SU(5) SUSY GUT.
This phase is provided by the complex phases 
in the SUSY breaking terms, for example, the phase
in the triple scalar coupling constant (A term). Since this phase
also induces electron and neutron EDMs, we have calculated the
T odd asymmetry in the $\mu^+ \rightarrow e^+ e^+ e^-$ taking
into account EDM constraints. 

\subsection{LFV branching ratios and asymmetries in the SU(5) and 
SO(10) SUSY GUT} 
We present results of our numerical calculation for SU(5)
and SO(10) SUSY GUT. We present following quantities
as contour plots in the SUSY parameter space.   
\begin{eqnarray}
\label{eq:data set}
&& \frac{B(\mu^{+}\rightarrow e^{+}\gamma)}{|\lambda_{\tau}|^{2}},~
\frac{B(\mu^{+}\rightarrow e^{+}e^{+}e^{-})}{|\lambda_{\tau}|^{2}},~
\frac{B(\mu^{+}\rightarrow e^{+}e^{+}e^{-})}
{B(\mu^{+}\rightarrow e^{+}\gamma)},~\nonumber \\
&& A(\mu^{+}\rightarrow e^{+}\gamma),~
A_{P_{1}}~,
A_{P_{2}},
A_T.
\end{eqnarray}
Here $\lambda_{\tau}$ is defined as
$\lambda_{\tau}=(V_R)_{23}(V_R)_{13}^{*}$ and $V_R$ is 
the right-handed lepton mixing matrix in the basis that the slepton 
mass matrix is diagonalized. In the minimal version of SUSY GUT
this mixing matrix is related to the CKM matrix at the GUT
scale, however in models which can explain realistic fermion mass
spectrum the simple relationship is lost. Therefore we treat
$\lambda_{\tau}$ as a free parameter and consider the quantities
listed above which are independent of this overall factor of 
the LFV amplitudes.

In Figs. 8 and 9 we show the above quantities  
in the plane of $m_{\tilde{e}_{R}}$ and $|A_{0}|$ for $\tan\beta = 3$,
$M_{2} = 300 $ GeV in the SU(5) SUSY GUT. 
\begin{figure} 
\begin{center}
\mbox{\psfig{figure=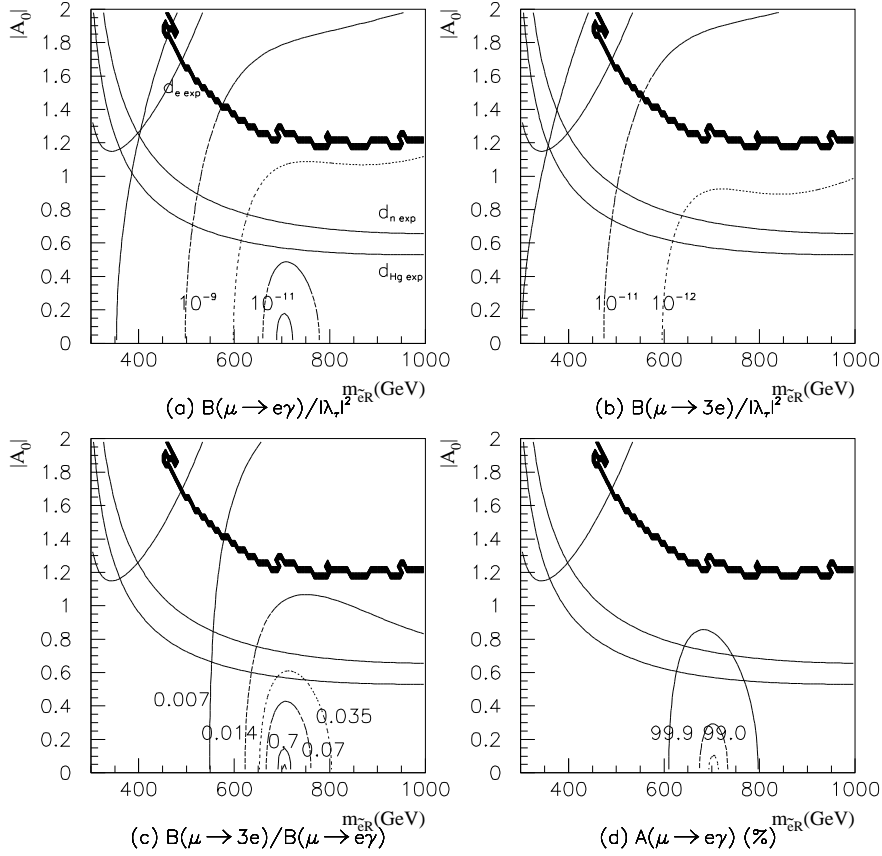 ,width=5in,angle=0}}
\end{center}
\caption{ The observables in the SU(5) model with the SUSY CP violating phases 
             in the $m_{\tilde{e}_R}$-$|A_0|$ plane.
             We fix the SUSY parameters as $\tan\beta = 3$,
             $M_2 = 300$ GeV, $\theta_{A_0}=\frac{\pi}{2}$
             and $\theta_{\mu}=0$
              and the top quark mass as 175 GeV.
             (a) Branching ratio for $\mu^{+} \rightarrow e^{+} \gamma$
              normalized by $|\lambda_\tau|^2$$\equiv$ 
             $|(V_R)_{23}(V_R)^\ast_{13}|^2$.
             (b) Branching ratio for $\mu^{+} \rightarrow e^{+}e^{+}e^{-}$
              normalized by $|\lambda_\tau|^2$. 
             (c) The ratio of two branching fractions 
             $\frac{B(\mu\rightarrow 3e)}
             {B(\mu\rightarrow e\gamma)}$.  
             (d) The P-odd asymmetry for $\mu^{+} \rightarrow e^{+} \gamma$.
             The experimental bounds from the electron, neutron
              and Hg EDMs are also shown in each figure.
             The left upper solid line corresponds to the electron EDM,
             the right upper solid line to the neutron EDM
             and the right lower solid line to the Hg EDM.
             The lower side of each bound is allowed by these experiments. 
             The upper side of the bold line is excluded by the EDM bounds
             even if we allow $\theta_{\mu}$ taking slightly different
              value from $0$. 
\label{fig:fig8}}
\end{figure} 
\begin{figure} 
\begin{center}
\mbox{\psfig{figure=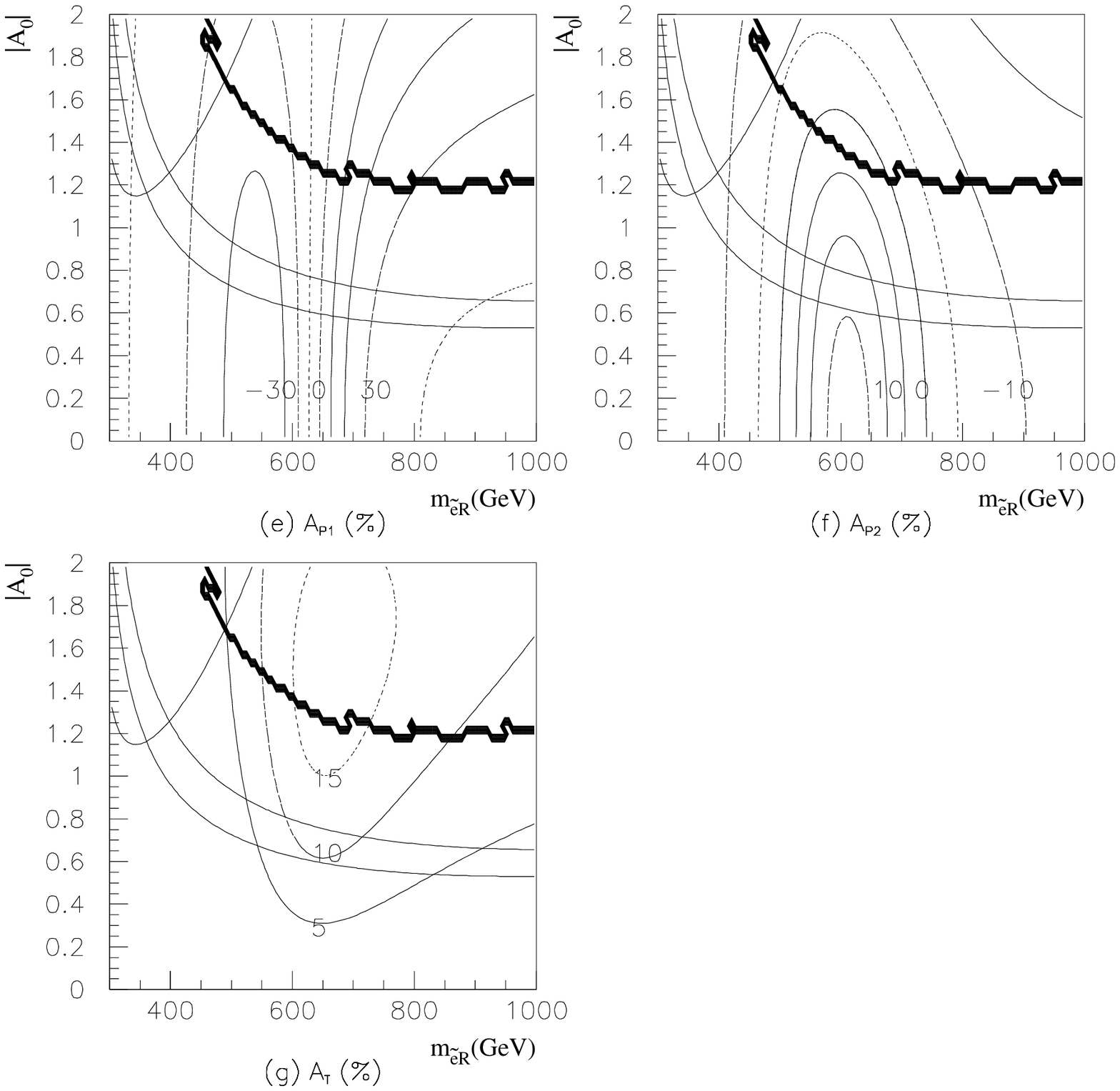 ,width=5in,angle=0}}
\end{center}
\caption{ Continued from the previous figure for the obsevables 
in the SU(5) model with the SUSY CP violating phases 
             in the $m_{\tilde{e}_R}$-$|A_0|$ plane.
             (e) The P-odd asymmetries $A_{P_1}$ 
             for $\mu^{+} \rightarrow e^{+}e^{+}e^{-}$.
             (f) The P-odd asymmetries $A_{P_2}$ 
             for $\mu^{+} \rightarrow e^{+}e^{+}e^{-}$. 
             (g) The T-odd asymmetry for 
               $\mu^{+} \rightarrow e^{+}e^{+}e^{-}$.
\label{fig:fig9}}
\end{figure} 
There are two independent
CP violating phases in the soft SUSY breaking paremeters which
we take the phases of A term ($\theta_{A_{0}}$) and the $\mu$ term 
($\theta_{\mu}$) at the Planck scale.  
In this figure we take  $\theta_{A_{0}}=\frac{\pi}{2}$ and
$\theta_{\mu}=0$. The experimental bounds from the electron, neutron
and Hg EDMs are also shown in each figure. Because these bounds
are sensitive to the small change of $\theta_{\mu}$, we also
show the parameter region which is not allowed 
even if we vary $\theta_{\mu}$ around $0$.  

We can see that for large parameter region the ratio of two branching
fraction is enhanced.  $A(\mu^{+}\rightarrow e^{+}\gamma)$ is close 
to 100\% except for small region where the almost exact cancellation 
occurs. The two P-odd asymmetries can be large and depend on
SUSY paremeter space. $A_{P_{1}}$ changes from $-30\%$ to $40\%$ 
and $A_{P_{2}}$ changes from $-10\%$ to $15\%$.
Within the allowed region of EDM constraints, 
the maximum value of $A_{T}$ is $15\%$. Note that in the SU(5) case
only $g_3$, $g_5$ and $A_L$ can be sizable because the LFV is induced 
in the right-handed slepton sector as long as we take not too large
$\tan\beta$ region. Using these asymmetries and branching ratios 
we can in principle determines these coupling constants up to an 
overall phase.

In Figs. 10 and 11 we show similar plots for SO(10) SUSY GUT case.
\begin{figure} 
\begin{center}
\mbox{\psfig{figure=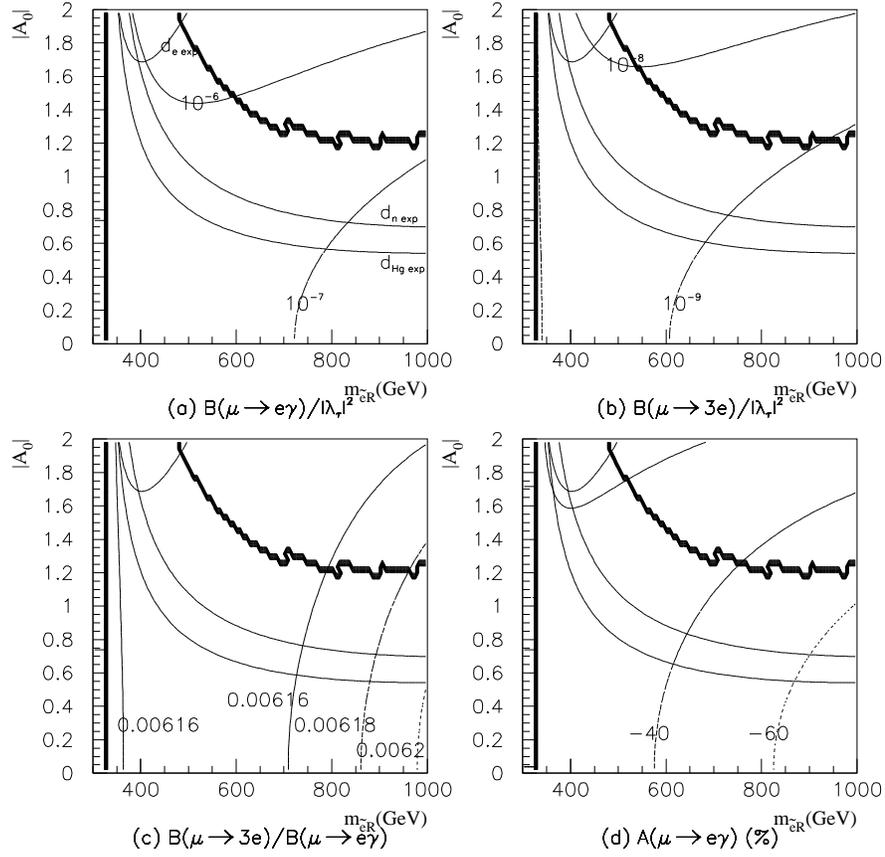 ,width=5in,angle=0}}
\end{center}
\caption{ The observables in the SO(10) model
              with the SUSY CP violating phase
              in $m_{\tilde{e}_R}$-$|A_0|$ plane. 
              The input parameters are same as 
              in Figs. 8 and 9..
              The small $m_{\tilde{e}_R}$ region bounded by the left
               bold line is not allowed in the minimal SUGRA model.
              The upper right bold line shows the bound from EDM 
              constraints set in the same way as in Figs. 8 and 9.
\label{fig:fig10}}
\end{figure} 
\begin{figure} 
\begin{center}
\mbox{\psfig{figure=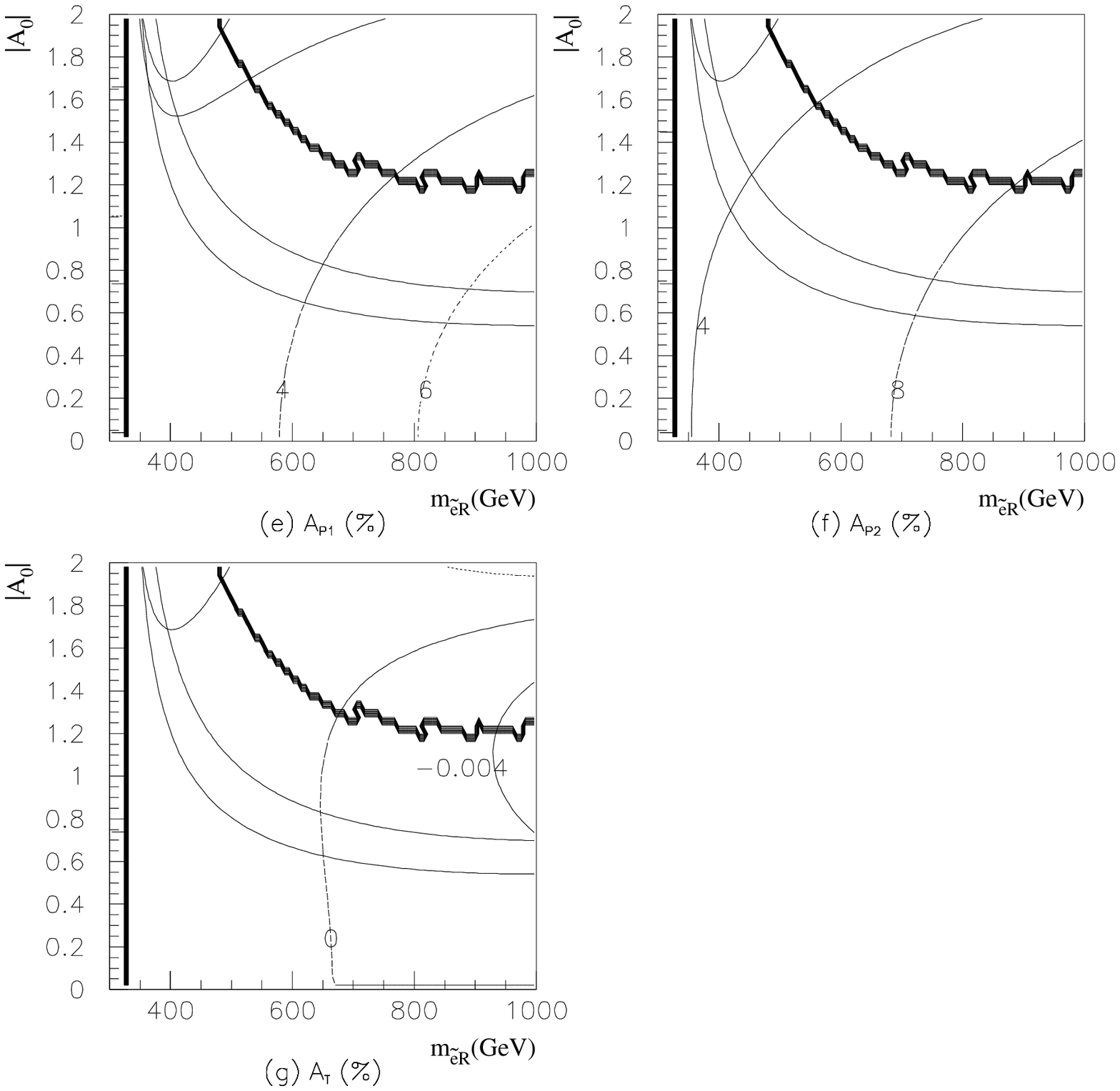 ,width=5in,angle=0}}
\end{center}
\caption{ Continued from the previous figure for the observables 
              in the SO(10) model
              with the SUSY CP violating phase
              in $m_{\tilde{e}_R}$-$|A_0|$ plane..
\label{fig:fig11}}
\end{figure} 
The $\mu^{+}\rightarrow e^{+}\gamma$ asymmetry 
$A(\mu^{+}\rightarrow e^{+}\gamma)$ varies from $-20\%$ to $-90\%$.
This is in contrast to the previous belief that $A_L$ and $A_R$ have 
a similar magnitude in this model.\cite{BHS} Although the diagram 
proportional to $m_{\tau}$ gives the same contribution to 
the $A_{L}$ and $A_{R}$, 
there is a chargino loop diagram which only contributes to $A_{R}$.
In spite of no $m_{\tau}$ enhancement, the contribution from the 
latter diagram can be comparable to that from the former one, 
especially when the slepton mass is larger than the chargino mass.
The P-odd asymmetries for $\mu^{+}\rightarrow e^{+}e^{+}e^{-}$ are 
proportional to $A(\mu^+\rightarrow e^+\gamma)$ for the SO(10) 
case. In fact we obtain approximate relations
\begin{eqnarray} 
A_{P_1} &\sim & -\frac{1}{10}A(\mu^{+}\rightarrow e^{+}\gamma),\\
A_{P_2} &\sim & -\frac{1}{6}A(\mu^{+}\rightarrow e^{+}\gamma).
\end{eqnarray}
For the branching ratio it is known that the following relation
is satisfied since the photon-penguin diagram give dominant
contributions. 
\begin{eqnarray} 
\frac{B(\mu^{+}\rightarrow e^{+}e^{+}e^{-})}
{B(\mu^{+}\rightarrow e^{+}\gamma)}
&\sim & 0.0062.
\end{eqnarray}
The above new relationship of the asymmetries also arises
since the $\mu^{+}\rightarrow e^{+}e^{+}e^{-}$ amplitude is
dominated by two photon-penguin amplitudes $(A_R, A_L)$.
As we see in the figure the T-odd asymmetry is very small
in the SO(10) model. This is also a consequence of the
photon penguin  dominance of the $\mu^{+}\rightarrow e^{+}e^{+}e^{-}$
amplitude because the four-fermion
amplitudes should be comparable in magnitude to the photon-penguin
amplitude to get sizable T-odd asymmetry.

We have investigated the branching ratios and asymmetries of 
two processes with different SUSY parameters. Qualitative features
are the same as in Figs. 8 - 11.  Results can be summarized
in Table 1 for SU(5) and SO(10) models.   
\begin{table}[h]
\begin{center}
\begin{tabular}{|c|c|c|} \hline
 & SU(5) SUSY GUT & SO(10) SUSY GUT \\ \hline
\hline
$A(\mu^{+}\rightarrow e^{+}\gamma)$ & $+100\%$ & $-20\%$~--~$-100\%$ 
\\ \hline
$\frac{B(\mu^{+}\rightarrow e^{+}e^{+}e^{-})}
{B(\mu^{+}\rightarrow e^{+}\gamma)}$
 & $0.007$~--~$O(1)$
 & constant~($\sim 0.0062$) \\ \hline
$A_{P_1}$
 & $-30\%$~--~$+40\%$
 & $\sim -\frac{1}{10}A(\mu^{+}\rightarrow e^{+}\gamma)$ \\ \hline
$A_{P_2}$
 & $-20\%$~--~$+20\%$
 & $\sim -\frac{1}{6}A(\mu^{+}\rightarrow e^{+}\gamma)$ \\ \hline
$|A_T|$
 & $\lsim 15\%$
 & $\lsim 0.01\%$ \\ \hline
\end{tabular}
\caption{Summary of branching ratios and asymmetries for
$\mu^{+}\rightarrow e^{+}\gamma$ and $\mu^{+}\rightarrow e^{+}e^{+}e^{-}$
processes in the SU(5) and SO(10) SUSY GUT.} 
\end{center}
\end{table}

\section{Conclusions}
In this talk we have discussed various processes sensitive
to loop diagrams of SUSY particles. For the quark sector
FCNC processes of $B$ and $K$ decays receive contributions
from flavor mixing in the squark sector. 
The muon anomalous magnetic moment is sensitive to 
the slepton loop diagram. LFV processes such as 
$\mu^{+}\rightarrow e^{+}\gamma$ and $\mu^{+}\rightarrow e^{+}e^{+}e^{-}$
are induced if there are flavor off-diagonal terms in the slepton mass 
matrices. 
 
We first updated the numerical analysis 
for FCNC processes in $B$ and $K$ decays 
and the muon anomalous magnetic moment in the supergravity model.
Taking account of the recent progress in the Higgs boson search, 
we showed that a small $\tanb$ region is almost excluded for $\tanb \lsim 2$.
The maximal deviation from the SM value in the \bb mixing is 12\% for
the minimal supergravity case and 30\% for the nonminimal case.
If the Higgs mass bound is raised to 105 GeV
the deviation is further reduced. We also calculate the SUSY contribution
to the muon anomalous magnetic moment. $\amu$ and $\brbsg$ show a strong 
correlation and $\amu$ becomes very large for a large $\tanb$ region.
We find that the SUSY contribution $\amu$ can be 
$(-30$ -- $+80) \times 10^{-10}$ for $\tanb = 10$ for the minimal
supergravity case. Along with the $\brbsg$ constraint, $\amu$ will soon
become a very important constraint on the parameter space in the
supergravity model. 

LFV processes in muon decay are particularly important
in the SUSY GUT because the flavor mixing in the quark sector
naturally induces LFV in the lepton sector through the interaction
at the GUT scale. This effect can be observed through the
renormalization effects on slepton mass matrices. 
The structure of the LFV effective Lagrangian at the muon scale
reflects nature of the LFV interaction at the high energy
scale.  

In order to see how we can distinguish various terms
in the effective Lagrangian 
we developed the model-independent formalism for 
$\mu^{+} \rightarrow e^{+} \gamma$ and 
$\mu^{+} \rightarrow e^{+}e^{+}e^{-}$ with polarized muon and defined 
convenient observables such as the P-odd and T-odd asymmetries.
Using explicit calculation based on the SU(5) and SO(10) SUSY GUT, 
we show that various combination of LFV coupling constants can be 
determined from the measurement of branching ratio and asymmetries.
In the SO(10) case the P-odd asymmetry in $\mu^{+}\rightarrow e^{+}\gamma$ 
varies from $-20\%$ to $-100\%$ whereas it is $+100\%$ for the SU(5) case.
We can define two kinds of P-odd asymmetries of $\mu^{+} 
\rightarrow e^{+}e^{+}e^{-}$. These asymmetries can be large and vary
over SUSY parameter space for the SU(5) case so that they 
can be used to determine effective coupling constants.
For the SO(10) case these two asymmetries are proportional to
the P-odd asymmetry in $\mu^{+}\rightarrow e^{+}\gamma$. 
We also calculated the T-odd asymmetry in the 
$\mu^{+}\rightarrow e^{+}e^{+}e^{-}$ process with the SUSY CP 
violating phases and compare it with the neutron, electron and Hg EDMs. 
The T-odd asymmetry can reach 15\% within the constraints of the EDMs
for the SU(5) case whereas it is very small for the SO(10) case.
Thus these quantities are useful to distinguish different models.

The experimental prospects for measuring these quantities depend on 
the branching ratio.
For the SO(10) model we expect 
the $\mu^{+}\rightarrow e^{+}\gamma$ branching ratio can 
be $10^{-12}$ when the $\lambda_{\tau}$ is given by the 
corresponding CKM matrix elements.
In such a case the $\mu^{+}\rightarrow e^{+}\gamma$ asymmetry 
can be measurable in an experiment with a 
sensitivity of $10^{-14}$ level.
For the SU(5) model, to get the $\mu^{+}\rightarrow e^{+}\gamma$ 
branching ratio of order $10^{-12}$ and the 
$\mu^{+}\rightarrow e^{+}e^{+}e^{-}$ 
branching ratio of $10^{-14}$, we have to assume $\lambda_{\tau}$ is 
larger than a several times $10^{-3}$.
If the branching ratio turns out to be as large, the 
$\mu^{+}\rightarrow e^{+}e^{+}e^{-}$ experiments with a sensitivity 
of $10^{-16}$ level could reveal various asymmetries.
Because various asymmetries are defined with respect to muon polarization, 
experimental searches for $\mu^{+}\rightarrow e^{+}\gamma$ and 
$\mu^{+} \rightarrow e^{+}e^{+}e^{-}$ with polarized muons are very 
important to uncover the nature of the LFV interactions.

\section*{Acknowledgments}
The work was supported in part by the Grant-in-Aid of the Ministry 
of Education, Science, Sports and Culture, Government of Japan (No.09640381),
Priority area ``Supersymmetry and Unified Theory of Elementary Particles'' 
(No. 707), and ``Physics of CP Violation'' (No.09246105).

\end{document}